\documentclass[reprint,amsmath,amssymb,prl]{revtex4-1}

\usepackage{graphicx}
\usepackage{dcolumn}
\usepackage{bm}
\usepackage[english]{babel}
\usepackage[utf8x]{inputenc}
\usepackage{amsmath}
\usepackage{color}

\newcommand{\figref}[1]{Figure~\ref{#1}}

\begin{document}
\title{Charge state dynamics of the nitrogen vacancy center in diamond under 1064 nm laser excitation}
\author{Peng Ji}
\author{M. V. Gurudev Dutt}
 \email{gdutt@pitt.edu}
\affiliation{%
 Department of Physics and Astronomy, University of Pittsburgh.
}%
\date{\today}

\begin{abstract}
The photophysics and charge state dynamics of the nitrogen vacancy (NV) center in diamond has been extensively investigated but is still not fully understood. In contrast to previous work, we find that NV$^{0}$ converts to NV$^{-}$ under excitation with low power near-infrared (1064 nm) light, resulting in \emph{increased} photoluminescence from the NV$^{-}$ state. We used a combination of spectral and time-resolved photoluminescence experiments and rate-equation modeling to conclude that NV$^{0}$ converts to NV$^{-}$ via absorption of 1064 nm photons from the valence band of diamond. We report fast quenching and recovery of the photoluminescence from \emph{both} charge states of the NV center under low power 1064 nm laser excitation, which has not been previously observed. We also find, using optically detected magnetic resonance experiments, that the charge transfer process mediated by the 1064 nm laser is spin-dependent.

\end{abstract}

\pacs{Valid PACS appear here}

\maketitle
\section{Introduction}

The negatively charged NV defect center (NV$^{-}$) in diamond has become prominent for applications in quantum information, nanoscale magnetic and electric field sensing, and fluorescent biological markers~\cite{Doherty13,Schirhagl14}. However, the photophysics of the NV$^{-}$ center, and the dynamics of charge transfer between NV$^{-}$ and NV$^{0}$ defect states, as well as the energy level structure and positions of the defect levels within the diamond bandgap, are still subjects of current research~\cite{Aslam13,Siyushev13,Doherty13}. Improved readout of the spin state via spin-to-charge conversion~\cite{Shields15} is crucially dependent on better understanding of the photophysics of the NV center.

Another exciting recent direction of research is the trapping of diamond nanocrystals for quantum opto-mechanics \cite{Yin13,Neukirch13,Hoang15}. Decrease of photoluminescence was observed from NV centers in the nanodiamonds that were optically trapped by an infra-red laser. In 2013, two groups observed fast photoluminescence  quenching and recovery under infrared excitation from nanodiamonds on a glass substrate \cite{Lai13,Geiselmann13}.  Explanations for these effects have included heating of diamond \cite{Hoang15}, multi-photon process \cite{Lai13}, charge state transfer \cite{Neukirch13} and also a dark state within the intrinsic optical transition of the NV center \cite{Geiselmann13}.  
In this work, we study the charge state dynamics of the NV center under 1064 nm laser excitation, through a combination of spectral and time-resolved photoluminescence and optically detected magnetic resonance experiments. We report four new findings: (i) the observation of \emph{enhanced} photoluminescence from the NV$^{-}$ charge state, (ii) fast modulation (quenching and recovery) of the photoluminescence ($\sim n s$) from \emph{both} NV$^{-}$ and NV$^{0}$ charge states, (iii) slow ($\sim \mu s$) charge transfer between NV$^{-}$ and NV$^{0}$ charge states under 1064 nm laser excitation, and (iv) evidence for spin dependent charge state dynamics under 1064 nm laser excitation. 

\section{EXPERIMENT}
\subsection*{Diamond samples}
The electronic grade bulk diamond sample (Element Six, $\left[ N \right] < 5$~ppb) was nitrogen implanted at 85 KV, with $10^{11}/\text{cm}^{2}$ dose, followed by annealing at 1000 $^\circ$C in forming gas ($\text{N}_{2}$ and $\text{H}_{2}$). This procedure creates a 20 nm NV center layer at about 100 nm below the diamond surface with estimated NV area density of $10/\mu \text{m}^{2}$. The 100 nm nanodiamonds (Adamas Nanotechnologies, Inc.) used in the experiment contain on average $\sim$ 500 NV centers in each nanodiamond (ND). These diamonds arrived suspended in de-ionized water and we diluted it with ethanol and deposit it onto clean silicon chips. \figref{setup} (c) shows the SEM image of these individual nanocrystals. We observed significant heterogeneity of the count rates from the ND, that could arise from the interplay of the crystal and NV orientation, laser polarization, charge state, and lifetime effects.  

\subsection*{Improving Thermal contact and Drift}
\begin{figure}[hbt]
 \includegraphics[width=6.5cm]{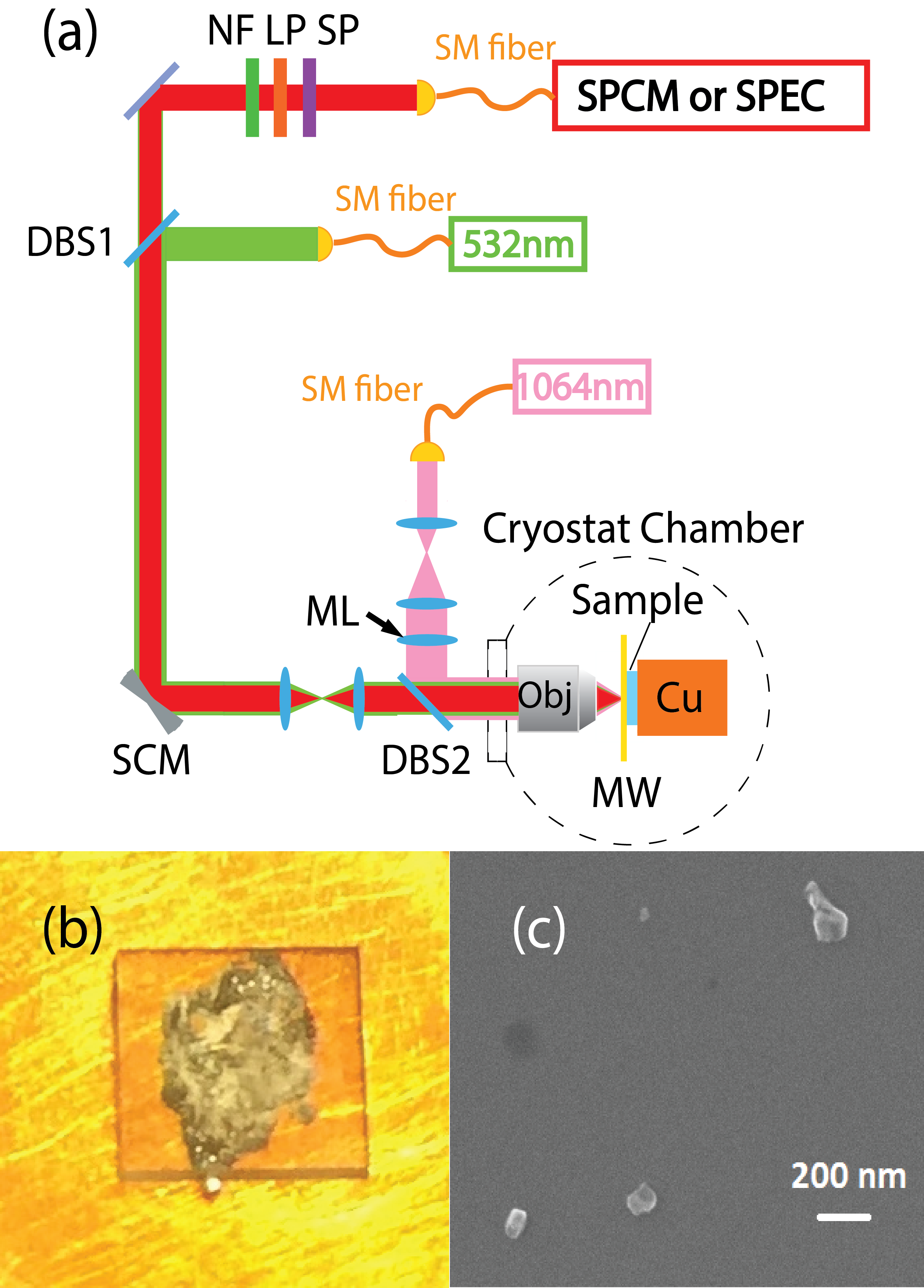}
\caption{\small{(a) Experimental confocal microscope integrated with 1064  nm CW laser. 532  nm laser, PL from NV centers and 1064  nm laser are separated by two dichroic beam splitters (DBS1 and DBS2). A moving lens (ML) in 1064  nm laser path helps to correct the chromatic aberration of the high NA Objective (Obj). Scanning mirror (SCM) helps to obtain 2D PL image. Samples are indium soldered on the cold finger made of copper (Cu) and microwaves are delivered with antenna (MW) to the sample. Notch 532  nm (NF) and different longpass (LP) and shortpass (SP) filters are applied for different collection windows (650  nm LP and 800  nm SP for NV$^{-}$ or 582/75  nm bandpass filter for NV$^{0}$). The PL was coupled into a single mode fiber (SM fiber) and received by Single Photon Counting Module (SPCM) or a spectrometer (SPEC) (b) Optical image of indium soldered bulk diamond on a copper plate (yellow), the indium solder (silver) can been seen through the bulk diamond, which is 3.5*3.5mm for its dimension (c) SEM image of 100 nm commercial nanodiamond deposited on a silicon chip. }}
\vspace{-7mm}
\label{setup}
\end{figure}
Our experimental setup is sketched in \figref{setup}(a). A confocal microscope is built for imaging the NV centers in diamond samples. We overlap the 532 nm excitation laser (Green) with a continuous wave (CW) 1064 nm laser (IR) and a long focal length lens was placed individually in the 1064 nm laser path to correct the chromatic aberration from the objective lens. The photoluminescence (PL) emitted by NV centers is collected with a high numerical aperture (NA) dry objective (Leica, Apo 0.9 NA). Different bandpass filters are applied for NV$^{-}$ or NV$^{0}$ as described below, and the PL is fiber-coupled into a Single Photon Counting Module (SPCM) for integrated counts or to a spectrometer. Microwaves are delivered through a 30 $\mu m$ in diameter gold plated tungsten antenna to the target NV centers to manipulate their ground spin states.  The objective and the samples can be enclosed in a cryostat chamber, which enables a low temperature environment.

Previous studies were carried out with very high peak-power pulsed 1064 nm excitation~\cite{Lai13} or moderate CW power~\cite{Geiselmann13} and did not mention the thermal drift of the sample caused by laser heating. However,  we found in our system that using a substrate with poor thermal conductivity such as cover glass, which was typically used with the earlier ND studies \cite{Lai13,Geiselmann13}, will result in thermal drift of the sample and reducing the collected PL counts on a time scale of ``seconds". We suspect this is possibly due to the thermal expansion of the substrate when tens of milliwatts of 1064 nm laser is intensely focused onto it. This drift can be corrected by re-optimizing the focus on the sample, but may interfere with the effects we are interested in. Therefore, we attached our sample to the cryostat cold finger (made of high purity oxygen-free copper), which serves as an excellent heat sink for the sample. The attachment between the sample(or silicon substrate for ND) and cold finger is by indium soldering.  We particularly choose indium soldering because this mechanical bond gives good thermal contact and can survive under low temperature. It also provides a lower PL background for a transparent sample compared to most optical adhesives. After indium soldering, under tens of milliwatts of 1064 nm laser illumination, we found that the counts were stable under IR excitation, and also found that re-optimizing the focus no longer improved the photoluminescence level, which suggests the thermal drift problem has been eliminated. \figref{setup}(b) shows the indium soldered bulk diamond sample on a copper plate (part of our cold finger).

\section{Results and Discussion}
\subsection{Steady state photochromism with IR excitation}
\begin{figure}[hbt]
\includegraphics[width=8cm]{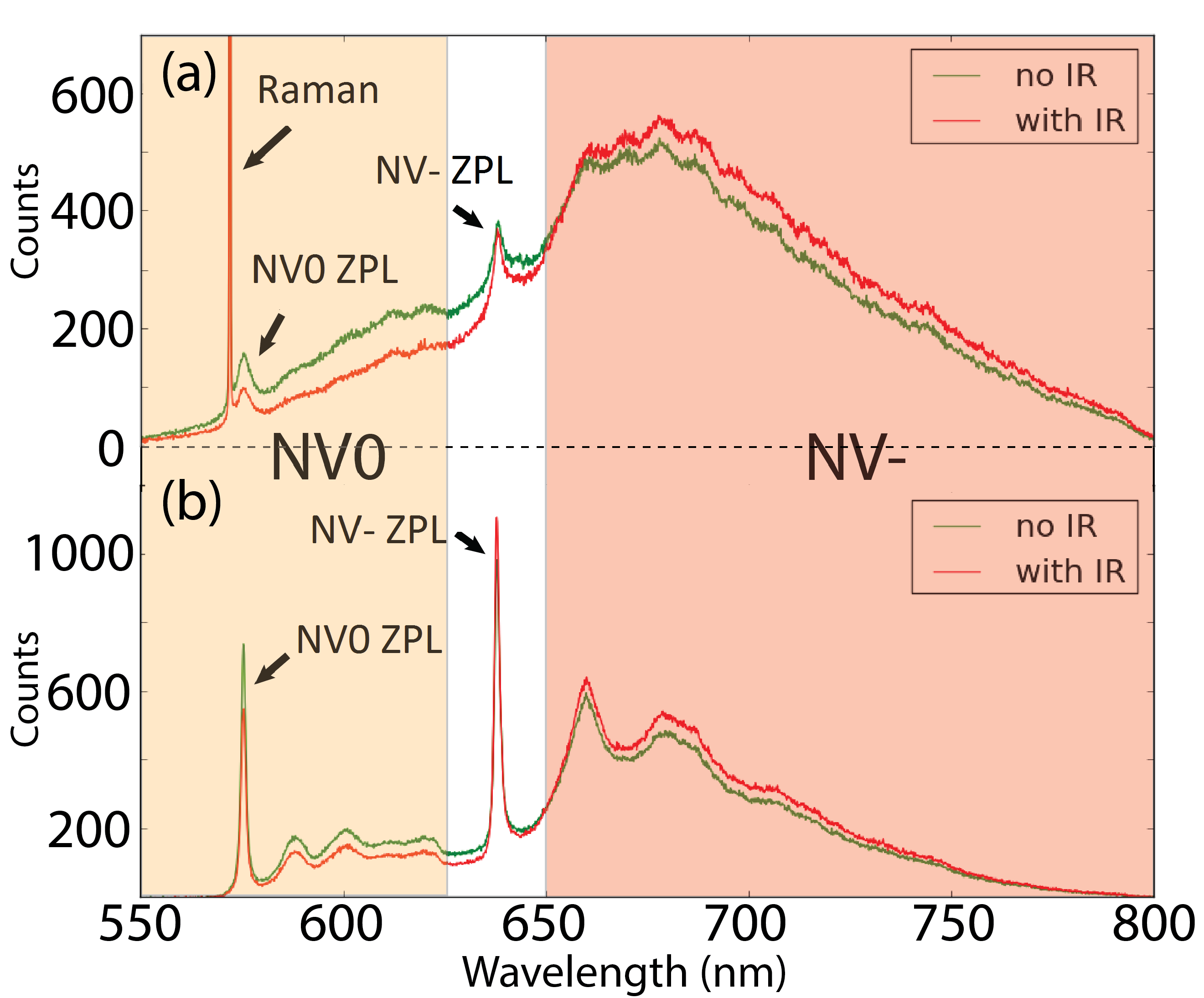}
\caption{\small{(a) Photochromism behavior of NV centers in bulk diamond at room temperature 295K under 1064  nm laser illumination (120 s exposure). ZPL of NV$^{0}$ and NV$^{-}$ are observed at 575 nm and 638nm; strong Raman peak is observed at 572.8 nm  (b) Photochromism behavior of NV centers in 100 nm diamond nanocrystal at 16K under 1064 nm laser illumination (30 s exposure). ZPL of NV$^{0}$ and NV$^{-}$ are observed at 575 nm and 637nm; no obvious Raman peak is observed at 572.8 nm. Shaded ranges indicate NV$^{0}$ (yellow) and NV$^{-}$ (red) PL collection window, shared with (a) and (b).}}
\label{spectrum}
\end{figure}

We find a marked change in the PL spectrum when exciting simultaneously with 532/1064 nm CW laser illumination. This phenomenon is robust at both room temperature and low temperature (\figref{spectrum}), and is visible in both bulk diamond and diamond nanocrystals. 

~The PL in the wavelength range from 550 nm to 625 nm has been identified as originating from NV$^{0}$, while it is usually accepted that the PL from the NV$^{-}$  phonon side band (PSB) dominates over that from NV$^{0}$ in the wavelength range from 650 nm to 800 nm \cite{Gaebel05}.  Since the PL change in \figref{spectrum} shows a decrease in the NV$^{0}$ band, while there is a corresponding increase at the same time in the NV$^{-}$ band, we believe the 1064 nm laser has the ability to induce the charge state flipping from NV$^{0}$ to NV$^{-}$.  For the bulk diamond sample at room temperature (295K) in \figref{spectrum}(a), we observed obvious PL decrease at the zero phonon line (ZPL) of  NV$^{0}$  centers (575 nm peak), while no marked PL increase was observed at the ZPL of NV$^{-}$ centers (638 nm peak at room temperature and 637nm at low temperature). This is because the PSB of NV$^{0}$ center slightly overlap with the NV$^{-}$ ZPL~\cite{Aslam13,Rondin10}. We observed similar PL change under low temperature (16K) from the 100 nm ND sample, as seen in \figref{spectrum}(b). One noticeable difference is the absence of a strong Raman line from the ND, which is expected in moving from bulk to~\mbox{nanocrystalline diamond}.

In the spectrally-integrated data that we present in the rest of the paper, we will refer to PL collected in the spectral window from 650-800 nm as arising from the NV$^{-}$ charge state, and to PL collected in the spectral window from 550 - 625 nm as arising from the NV$^{0}$ charge state. This is in good agreement with previous observations, as well as with our data shown in \figref{spectrum}.

\subsection{Time resolved photoluminescence}
Nanodiamonds that we purchased commercially have large heterogeneity and therefore we focused on the bulk diamond sample to further study this 1064 nm laser induced effect. 
\begin{figure}[hbt]
 \includegraphics[width=8cm]{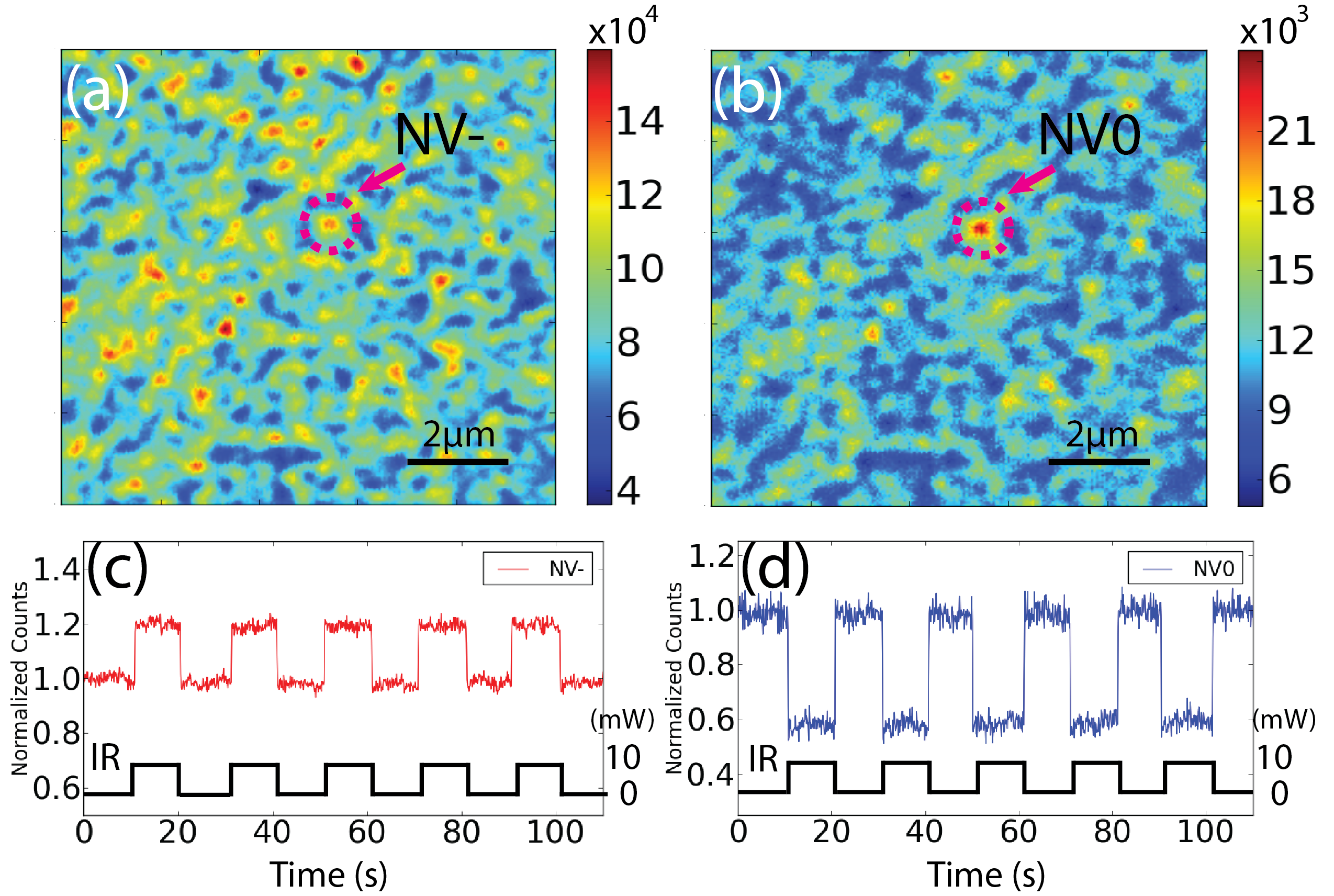}
\caption{\small{Confocal images of ensemble of NV centers in bulk diamond sample at room temperature under 0.25mW Green excitation (a) NV$^{-}$ PL data (b) NV$^{0}$ PL data (c) Time trace of the NV$^{-}$ PL counts, under IR gating (10 mW), normalized to when IR is off (d)  Time trace of the NV$^{0}$ PL counts, under IR gating (10 mW), normalized to when IR is off.}}
\label{confocal}
\end{figure}
Firstly, the confocal images of NV centers in the bulk diamond were taken in \figref{confocal}. We chose randomly a location where both charge states gives sufficient counts rate as shown in \figref{confocal}(a) and (b). Then, we turn on and off the IR laser while green excitation is on all the time. The data shown in \figref{confocal} (c) and (d) are obtained by integrating the counts in the spectral window corresponding to NV$^{-}$ and NV$^{0}$ respectively.  
The increase (decrease) of the counts in the two spectral regions, modulated by the IR laser, is robust and is the opposite of previous experiments in ND that found only decreased photoluminescence of the NV$^{-}$ state under IR excitation~\cite{Lai13,Geiselmann13}.

\begin{figure}[hbt]
\centering
 \includegraphics[width=8cm]{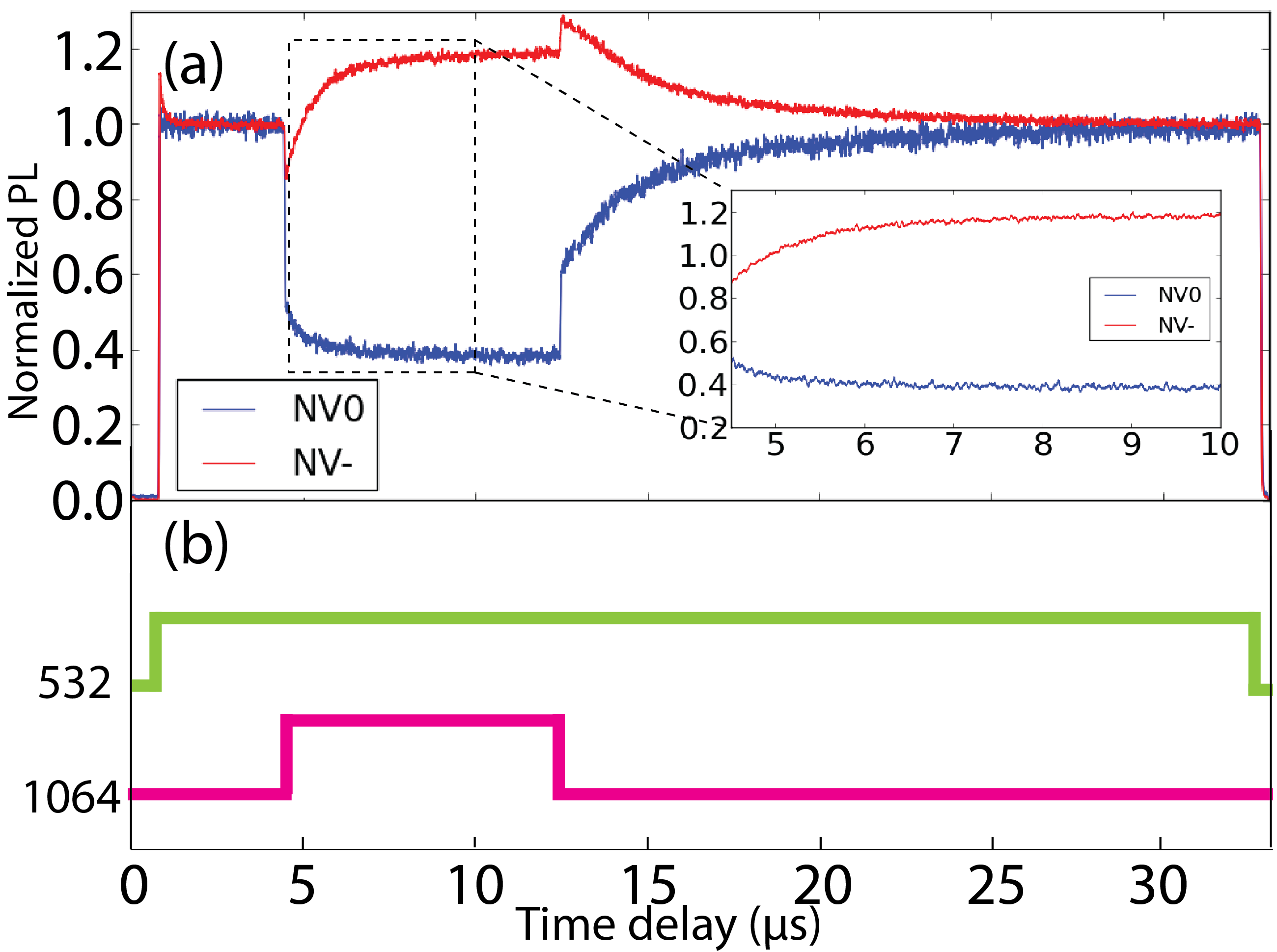}
\caption{\small{(a) PL versus time (Savitzky–Golay filter applied \cite{savitzky1964}) while fast switching on and off 1064 nm laser applied by AOMs; inner figure: Zoom in of the ``slow effect" (b) Optical pulsed sequence used in the experiments for result (a)}}
\label{30us}
\end{figure}

We modulated both green and IR lasers on sub-microsecond timescales to time-resolve the PL from NV$^{-}$ and NV$^{0}$. We used an acousto-optic modulator (AOM) in our experiments to realize fast switching of both 532 nm and 1064 nm lasers. The response times of both AOMs are $\sim 50$ ns. The first experimental pulsed sequence and result are shown in \figref{30us}. The PL is collected for both states under same experimental condition and then normalized to the steady state PL separately. 

As shown in \figref{30us}, after the 1064 nm laser was on, we first observed a  fast quench of PL counts for both charge states. It was followed by a slow increase for the  NV$^{-}$ state and a slow decrease for  NV$^{0}$ state separately. Neglecting the fast decrease for the moment, we call the latter effect the ``slow effect". Below, we have developed a simple theoretical model (\figref{model}) based on NV$^{0} \leftrightarrow$ NV$^{-}$ charge state transfer which explains quite well the slow effect.

\subsection*{Theoretical model}
\begin{figure}[hbt]
\centering
 \includegraphics[width=8cm]{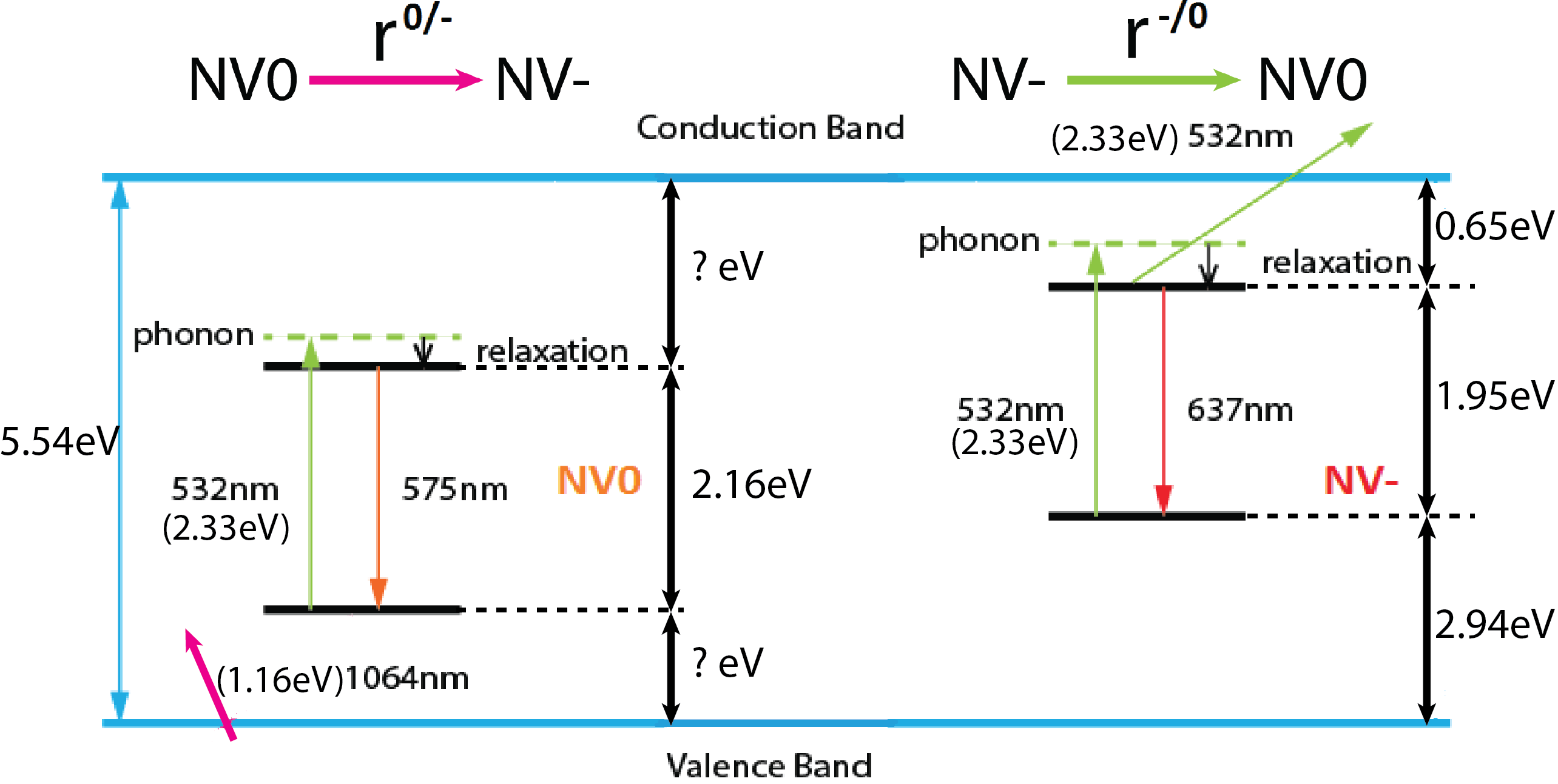}
\caption{\small{Theoretical model for NV charge state flipping. Left: our postulated model that NV$^{0}$ is excited first by 532 nm laser (green arrow), then it captures an electron promoted by 1064 nm laser form the valence band (pink arrow). This two-step procedure converts  NV$^{0}$ to  NV$^{-}$. Right: accepted photo-ionization two-step procedure induced by 532 nm laser (green arrows indicate the two steps), which converts NV$^{-}$ to  NV$^{0}$~\cite{Aslam13} through the conduction band.}}
\label{model}
\end{figure}

The relative location of the NV$^{-}$ charge state to the conduction and valence band of diamond has been established as shown in \figref{model} (following Ref.\cite{Aslam13}). The current understanding is that the photo-ionization from NV$^{-}$ to NV$^{0}$ is actually a two-step procedure: one laser photon with energy $> 1.95$ eV excites the ground state NV$^{-}$ to the excited state, followed by another laser photon ionizes it through diamond conduction band, resulting in the NV$^{0}$ charge state~\cite{Manson05,Waldherr11,Beha12,Aslam13}.  

Meanwhile, the position of NV$^{0}$ energy levels relative to the conduction and valence bands is not well known and the source of the electron captured by NV$^{0}$ to convert it back to NV$^{-}$ is still under debate. One possible electron donor is the valence band of diamond. Similar to the ionization process, the NV$^{0}$ $\rightarrow$ NV$^{-}$ process was also suggested to be a two-step procedure, which requires one laser photon with energy $> 2.15$ eV to excite NV$^{0}$ to its excited state, followed by another photon that promotes an electron from the valence band, which is captured by NV$^{0}$ to form NV$^{-}$ charge state \cite{Siyushev13,Beha12,Aslam13}. This explanation met trouble when experimental detection of NV$^{0}$ $\rightarrow$ NV$^{-}$ process occurred under 593 nm ($2.1$ eV) laser illumination, in which case the NV$^{0}$ charge state should not be excited \cite{Aslam13}. The substitutional nitrogen impurity (P1 center) nearby was suggested as another possible source of the electron \cite{Robledo10,Aslam13}. In this one-step procedure, NV$^{0}$ captures the optically excited electron from nearby P1 donor through conduction band, and is converted to NV$^{-}$. However, since the P1 donor level is measured to lie $1.7\sim 2.2$~eV below the conduction band~\cite{Walker1979},the 1064 nm laser (energy of 1.165 eV) should not be able to trigger this one-step transformation. Ref.\cite{Robledo10,Aslam13} also indicate thresholds as 1.95 eV or 2.03 eV separately to trigger this one-step procedure based on their data.   

\begin{figure}[hbt]
\centering
 \includegraphics[width=8cm]{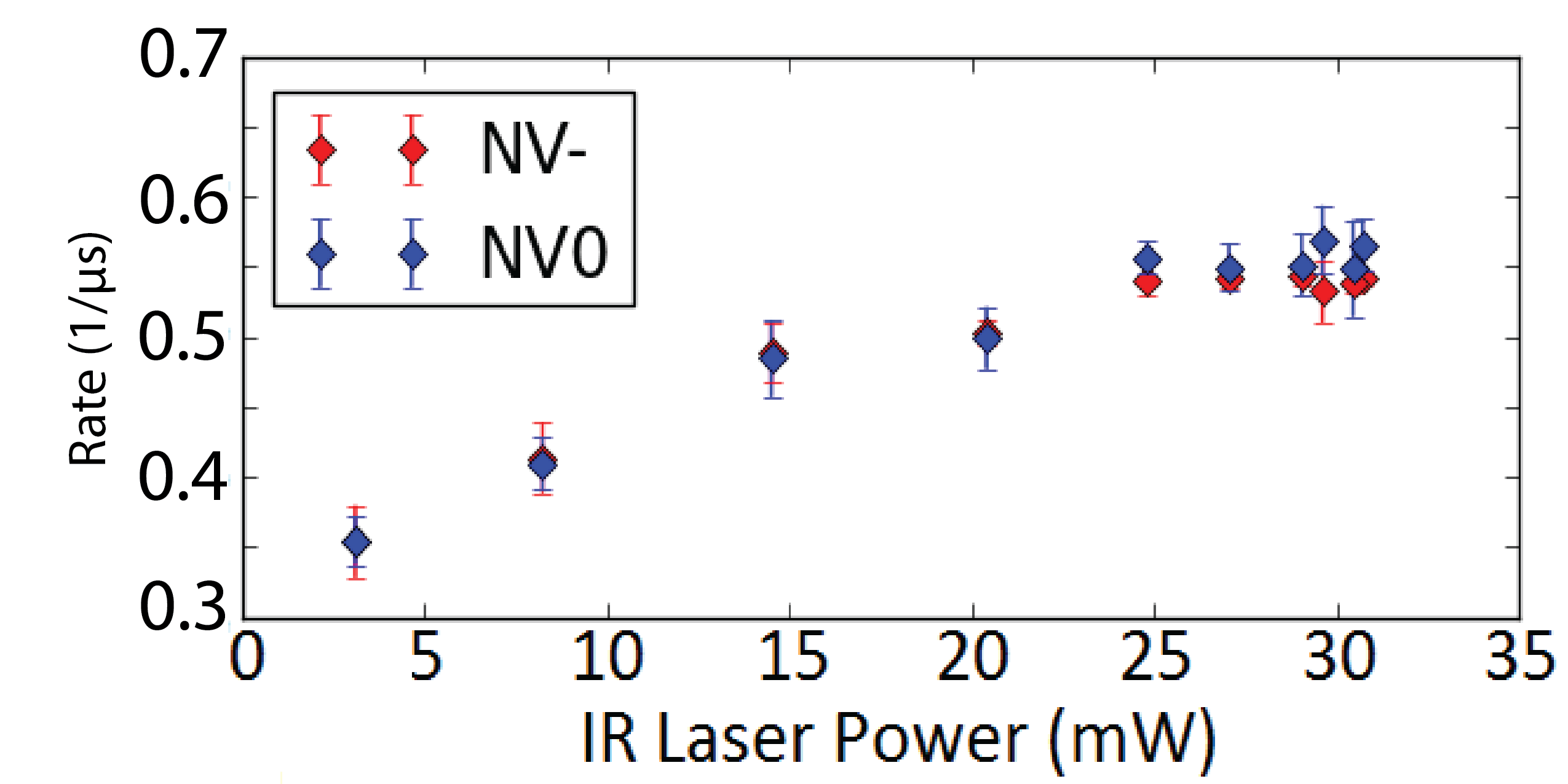}
\caption{\small{Dependence of the charge flipping rate $r=r^{-/0}+r^{0/-}$ on the 1064 nm laser power. Green laser excitation is fixed at 0.3 mW. }}
\label{rate}
\end{figure}

Therefore, we postulate an extension for the two-step NV$^{0}$ $\rightarrow$ NV$^{-}$ model that l064 nm laser may provide sufficient photon energy to complete the second step, where a 1064 nm laser-excited electron from valence band got captured by a NV$^{0}$ to form a NV$^{-}$. This extended two-step process competes with the well studied ionization procedure caused by 532 nm laser (converts NV$^{-}$ to NV$^{0}$ ), forming a new balance of charge states population. 
We use a simplified two-state (NV$^{-}$ and NV$^{0}$) model to describe the dynamics of charge state flipping. The populations of the charge states can be described as follows:
\begin{align}
  \dfrac{dP^{-}}{dt} = -r^{-/0}*P^{-} + r^{0/-}*P^{0}\\
  \dfrac{dP^{0}}{dt} = +r^{-/0}*P^{-} - r^{0/-}*P^{0} 
\end{align}
where $P^{i} (t)$ (i = - or 0) are the population of two charge states respectively as a function of time, $r^{-/0}$ is charge flipping rate from negative to neutral state while $r^{0/-}$ is the opposite. The dynamic solution indicates an exponential dependence of populations versus time:
\begin{align}
  P^{i}(t) \sim \exp (-(r^{-/0}+r^{0/-})*t)
\end{align}
We used this result to fit the ``slow effect" shown in \figref{30us}(a) and obtain the dependence of $r=r^{-/0}+r^{0/-}$ on 1064 nm laser power , which is shown in \figref{rate}.

From \figref{rate}, firstly, it is clear that the rates of the slow changes in PL collected from NV$^{-}$ and NV$^{0}$ are synchronized reasonably well, which indicates again that charge state flipping argument is correct. Secondly, the charge flipping time constant is in microseconds time scale, which is much slower than the internal optical transitions of NV center\cite{Doherty13}. This time scale also agrees with the conclusions in Ref.~\cite{Aslam13} under IR illumination power we used in experiments. Finally, we found a linear relation between IR power and the sum of flipping rates in \figref{rate} when IR power is moderate ($< $15 mW). We believe the saturation behavior after that is limited by the specific process that happens during electron promotion, such as saturation from weak green excitation and optical pumping.

\begin{figure}[hbt]
\centering                                                               \includegraphics[width=8cm]{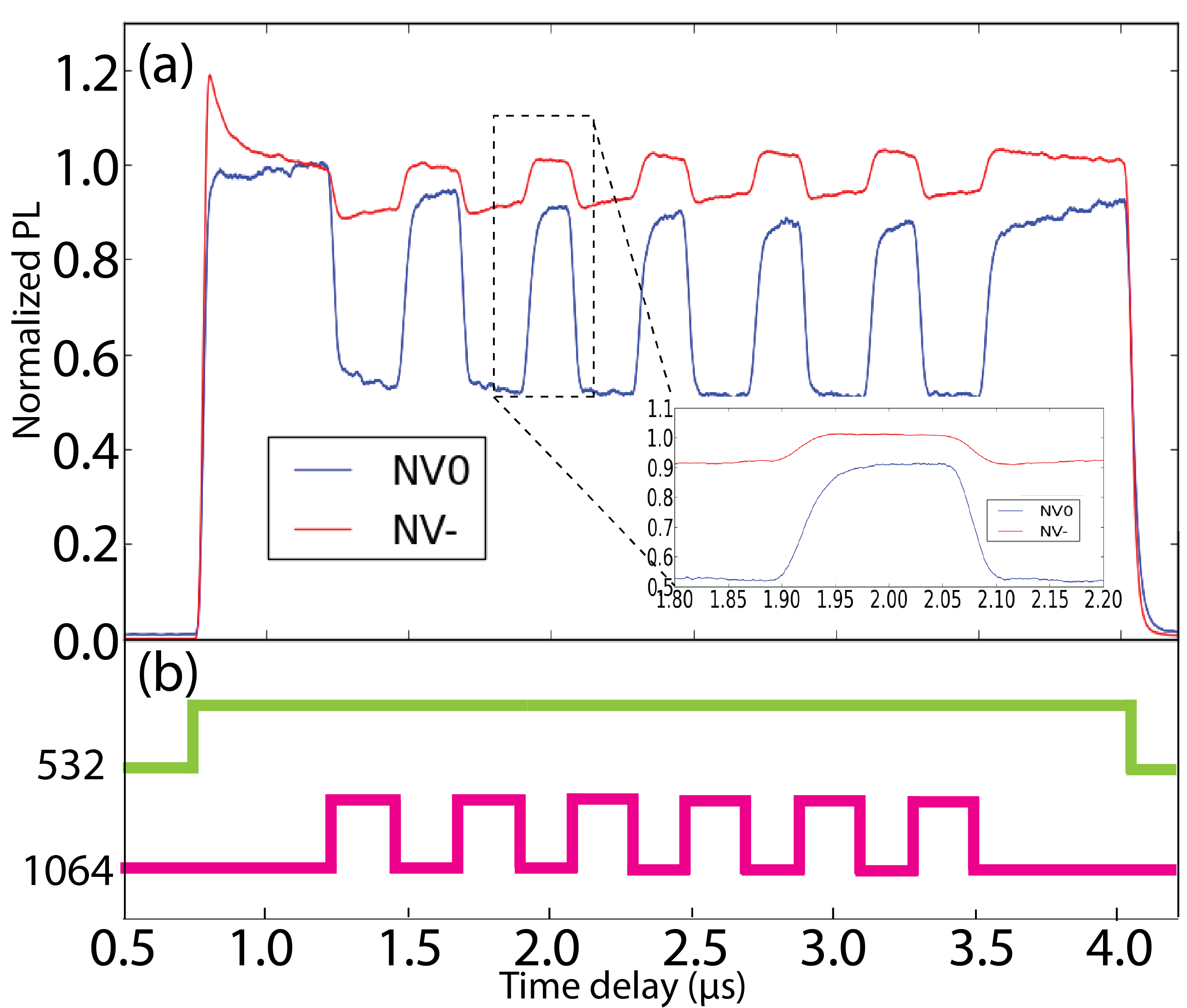}
\caption{\small{(a) PL versus time (Savitzky–Golay filter applied \cite{savitzky1964}) under fast modulation to examine the rapid quenching behavior; inner figure: Zoom in of the fast edges of PL counts (b) Optical pulse sequence used for the results in (a)}}
\label{200ns}
\end{figure}

\subsection*{Fast modulation of PL from both charge states}
We now return to the rapid quenching and recovery of the PL which we observed for both charge states in \figref{30us}. To gain further insight, we applied 1064 nm laser pulses of width 200 ns repeated several times, as shown in \figref{200ns}. As seen in the data, the fast decrease and subsequent recovery for both charge states is definitely limited by response time of our AOM ($\sim 50$ ns). 

The first observation of 1064 nm laser induced PL fast quenching for both charge states was  reported by Ref.~\cite{Lai13}. In their experiments, a high peak power 1064 nm pulsed laser (4.8 MHz repetition rate, 16 ps pulse duration) was focused by 1.4 NA oil immersion objective onto ND samples. We estimated that the focused peak power density $\sim 400 $ W/$\mu m^2$ from the parameters given in that work. The authors attributed the fast quenching time of hundreds of picoseconds to multiple-photon (MP) related thermal induced change in NV photophysics parameters. However, in Ref.~\cite{Geiselmann13}, a similar quench was reported on single NV$^{-}$ centers in diamond nanocrystal under very different conditions using a CW IR laser at 1064 nm. In that work, the laser was focused onto ND deposited on cover glass by 1.2 NA oil immersion objective, and we estimated that the focused power density $\sim 65$ mW/$\mu m^2$ from their parameters. They argued their laser was not powerful enough to trigger MP procedure. The explanation of their data was postulated to be a previously unreported dark state in the optical transitions of the NV$^{-}$ center. However, they did not report that this phenomenon also exists and is even more pronounced for NV$^{0}$ charge state as we show in \figref{200ns}.

Based on the previous results, and our experimental conditions, we believe our data also does not originate from any thermal heating or MP processes. Firstly, we are using a CW laser which is being modulated on much slower timescales in comparison to the pulsed laser in Ref.\cite{Lai13}; secondly, because of the 0.9 NA dry objective we use compared to Ref.\cite{Geiselmann13}, our laser focused intensity is even lower ($\sim 20$ mW/$\mu m^2$ maximum). We also provided a much better thermal contact to a good heat sink material, and in comparison to ND, we used bulk diamond which has extremely high thermal conductivity. Ref.\cite{Lai13} also claimed they did not see obvious effect by using CW 1064 nm laser, which is the excitation that we use. Therefore we believe that the fast quenching is not because of the heating or MP processes caused by the 1064 nm laser. 

In our model, when low green illumination is used, the excited state of both charge states is populated only weakly. The first possibility is that the IR laser causes transitions from the ground state to one of the metastable electronic levels. However, while the NV$^{-}$ center has a metastable singlet transition ($^1A_1$ to $^1E$) at $1042$~nm, it is unlikely that a transition will occur at the low excitation powers we use, and further the typical 300~ns singlet relaxation time is much longer than the fast recovery we see. Another possibility is that the IR photon causes a fast deformation of the lattice away from the $C_{3v}$ symmetry via vibronic coupling, resulting in dressing of the electronic levels by a Jahn-Teller like splitting, and thereby decreasing the dipole matrix elements and the photoluminescence from the electronic levels. Stimulated emissions from the NV excited state to its ground state vibronic side band is another explanation for the fast quenching. A recent work observed a similar quenching under the stimulating wavelength of $\sim$ 700 nm \cite{Jeske16} (from a pulsed source), which is the peak of the NV emission PSB. They argued this fast effect as stimulated emission because they observed PL reductions at the rest of wavelengths (550 nm to 650 nm). However, in our case, considering the extremely weak spontaneous PSB emission at 1064 nm from both charge states and also the weak laser power density we used compared to their work, direct observation of strongly enhanced 1064 nm emission will be needed before we can attribute the fast effect to stimulated emission. Finally, it is possible that the IR photon is causing a fast local change in the electric field around the NV centers, for instance through optical rectification (a $\chi^{2}$ process) or altering the band-bending at the surface of the diamond which is $\sim 100$~nm away, which also will affect the dipole matrix elements. We are not aware of previous literature where such effects have been explicitly identified. The origin of this fast modulation is clearly an important subject for future research.  

\subsection{Optically detected magnetic resonance (ODMR) under 1064 nm laser Illumination}

We found that the charge state flipping of NV centers, especially with IR illumination, has been rarely studied with optically detected magnetic resonance (ODMR) before. Therefore, we decided to carry out ODMR at the confocal spots where we had already observed charge flipping (sample shown in \figref{confocal}).

\begin{figure}[hbt]
\centering
\includegraphics[width=8cm]{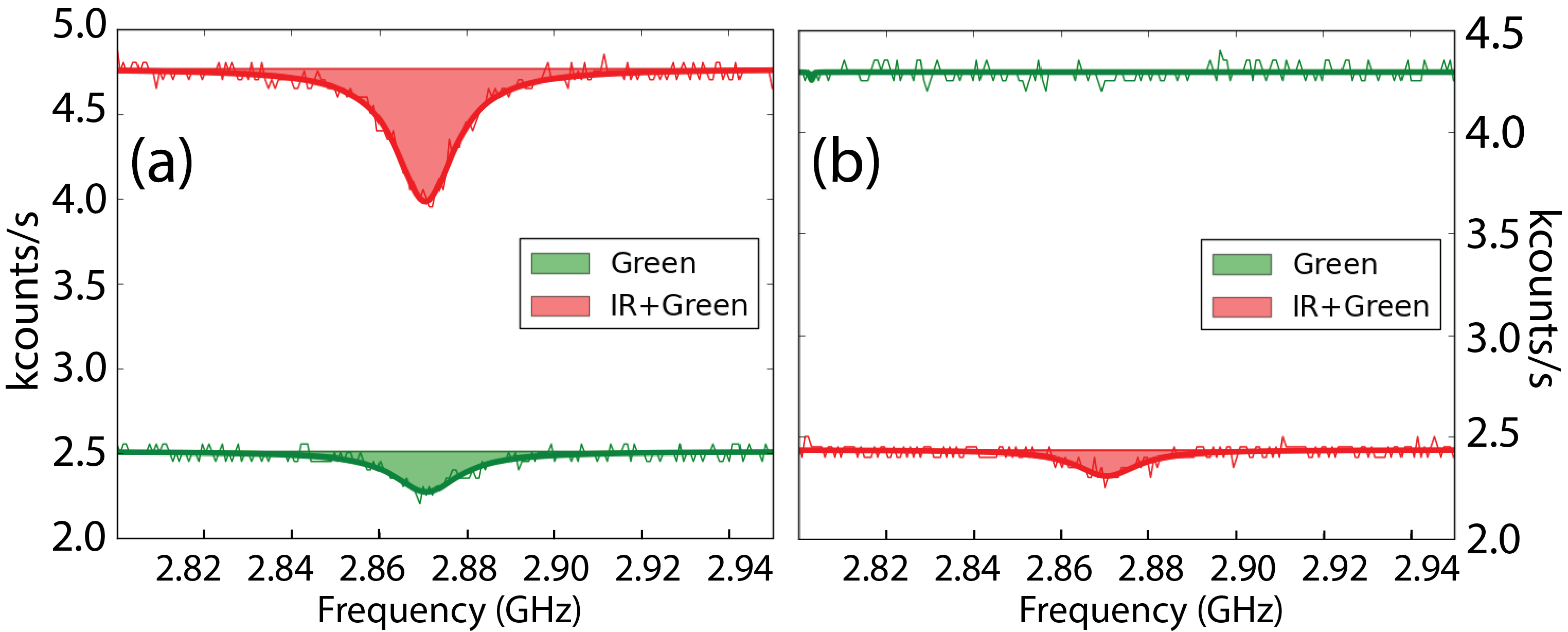}
\caption{\small{ODMR signals comparison with IR and no IR for (a) NV$^{-}$ charge state (b) NV$^{0}$ charge state}}
\label{ESR}
\end{figure}

The first interesting finding is 1064 nm laser enhances NV$^{-}$  ODMR contrast under low 532 nm laser excitation power. In this experiment, we used 30 mW 1064 nm laser power and 0.013 mW 532 nm laser power for excitation, which was carefully optimized to improve the contrast. The CW ODMR experimental result for NV$^{-}$ collection window (650-800 nm) is shown in \figref{ESR}(a). It is very clear that the participation of 1064 nm laser increased the overall PL counts by a factor of 1.9; meanwhile, the ODMR contrast at 2.87 GHz has also been enhanced by a factor of 1.7. This enhancement effect is robust through the sample, although the amount of this increase is different from one location to another. 
Because the four orientations of NV centers within the laser spot could be randomly distributed, the effective laser power used in charge flipping dynamics could be different from one location to another. Therefore, it is reasonable that the final populations are different from one location to another. 

The thermally-induced ODMR peak shift has been well studied~\cite{Acosta10,Kucsko13}. We did not see any shift of the ODMR peak away from 2.87 GHz during repetitive tests, within our signal to noise limit. The absence of this shift indicates again that our observations are unlikely to be due to temperature shifts.

Another interesting finding appears when we switch to NV$^{0}$ collection window (550-625 nm). We set 20 mW 1064 nm laser with 0.1 mW excitation 532 nm laser for optimizing the effect. The overall background counts decrease, opposite to NV$^{-}$ window, consistent to the result in \figref{spectrum} (a) and (b). However, we observe an unexpected ODMR signal around 2.87GHz from the NV$^{0}$ collection window in \figref{ESR}(b) when the IR laser excitation is used. 

Since we have not applied magnetic field during this experiment, and due to spin 1/2 nature of NV$^{0}$,  we postulated that the charge state flipping procedure under IR excitation is spin dependent. It tends to flip more NV$^{0}$ to NV$^{-}$ when NV$^{-}$ is populated in the $\lvert m_s = \pm 1 \rangle  $ state. 

\begin{figure}[hbt]
\centering
 \includegraphics[width=8cm]{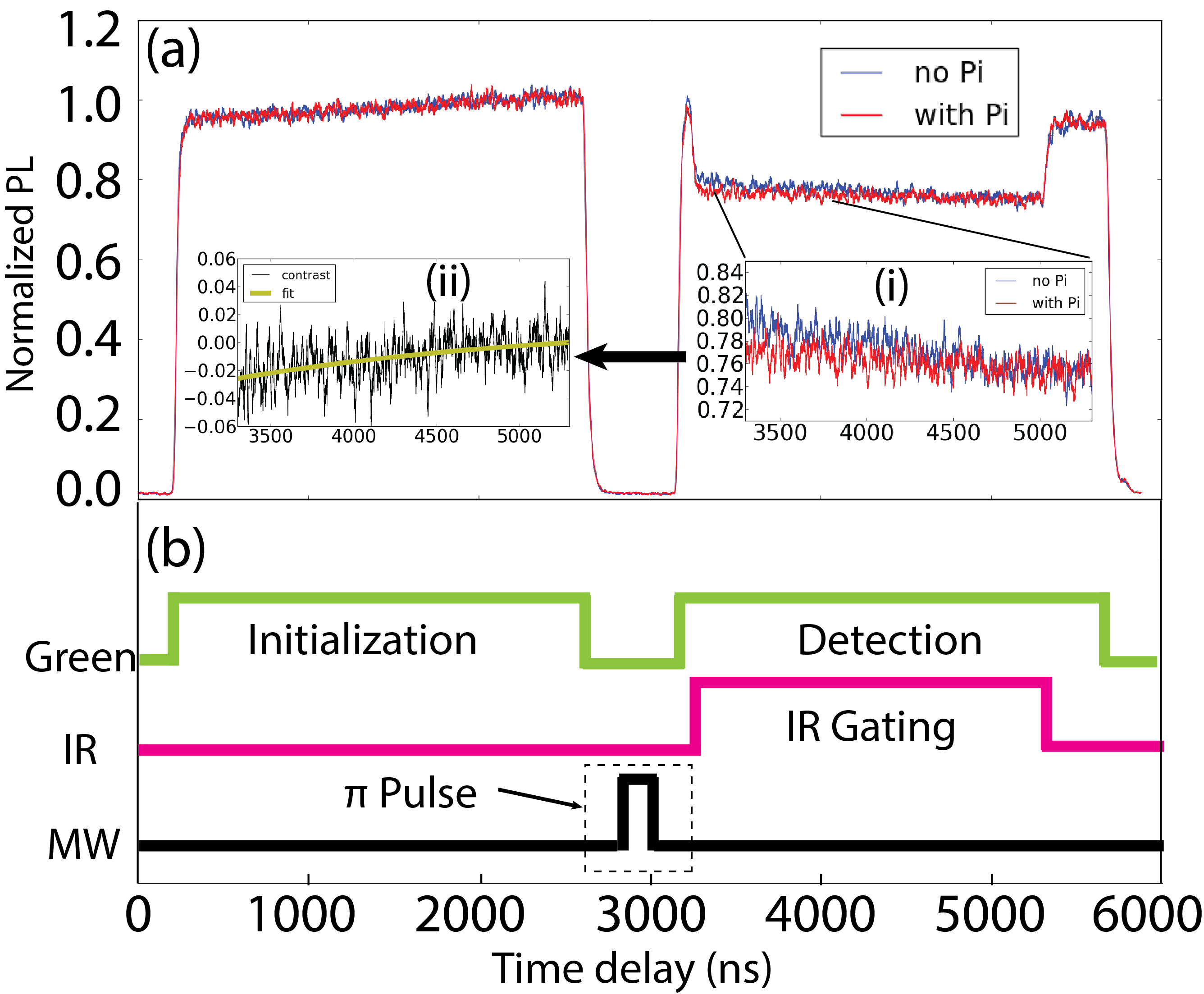}
\caption{\small{(a) Spin dependence of NV$^{0}$  PL counts (Savitzky–Golay filter applied \cite{savitzky1964}, normalized to steady state) through charge state flipping after switching on 1064 nm laser (i) Zoom in of PL contrast (ii) Exponentially fit of the PL contrast decay time (b) Corresponding pulsed sequence: start with green laser initialization pulse for 2300 ns, followed wtih a 2.87GHz $\pi$ pulse to flip the spin after green was off. A detection green pulse  was applied 100 ns after the Pi pulse and IR gating pulse was 50 ns after that. }}
\label{pulse}
\end{figure}

To further examine this effect, we attempted a pulsed experiment which is designed to show the spin dependence property of NV$^{0}$ $\rightarrow$ NV$^{-}$  charge state dynamics. The pulsed sequences are shown in \figref{pulse}(b). We start the sequence with initialization green laser pulse, followed with a microwave $\pi$ pulse after switching off the green laser. We independently can measure Rabi oscillations to calibrate our $\pi$ pulse (data not shown) by using just the green laser and following standard time-resolved counting techniques for NV center \cite{Jelezko04,Childress06}. Thus, this procedure populates the NV$^{-}$ mostly to its $\lvert m_s = \pm 1 \rangle $ state with the $\pi$ pulse. Then the green laser was switched on again 100 ns after the $\pi$ pulse and the IR laser was switched on 50 ns after the green laser. We try to compare the behavior after IR laser was on for the cases when microwave $\pi$ pulse is switched on or not. Normalized results are shown in \figref{pulse}(a). As we expected, mapping NV$^{-}$ to its spin-1 state by the $\pi$ pulse seems to enhance the charge state flipping NV$^{0} \rightarrow $ NV$^{-}$ rate and therefore less PL counts are collected from NV$^{0}$ window. A zoom in figure (i) shows the PL difference of the two cases. Although the contrast is small, this difference is robust from one run to another. We exponentially fitted the PL contrast in figure (ii), which indicates a 1900 ns decay time before the contrast disappeared. 

This decay time may seem long considering the measured $\sim 300$ ns spin polarization time from the $m_s = \pm 1$ to $m_s = 0$ state of the NV$^{-}$ center. However, we note that given the excitation power of the green laser which is well below saturation, the rate of spin-polarization by optical pumping will be decreased. While we have not carried out detailed calculations, it seems reasonable given the data that the contrast and the time-scale shown in \figref{pulse} arises from optically mediated spin polarization under our experimental conditions. The main feature we are reporting in this paper is that there appears to be a spin-dependent component to the charge transfer, as confirmed by the fact that we see ODMR signal when collecting PL from the NV$^{0}$ window under IR excitation, and the time-resolved experiment with and without a microwave $\pi$ pulse shows a contrast when the IR laser is turned on.

\subsection{Conclusion}
In conclusion, we report the observation of 1064 nm laser induced charge state flipping from NV$^{0}$ to NV$^{-}$ in small ensembles of NV centers, under a wide variety of samples and experimental conditions. A faster (50 ns, limited by our AOMs) PL quenching and recovery effect for both charge states has been observed but is still not fully understood. We analyzed the charge state flipping model, which indicates 1064 nm laser photon could be sufficient to promote an electron from valence band and captured by NV$^{0}$ to convert to NV$^{-}$ state, and found good agreement with the slow ($\sim \mu$s) PL dynamics. We also found preliminary evidence that the charge state flipping induced by the 1064 nm laser is spin dependent by carrying out ODMR experiments.

Our work is a first step in trying to unravel the photophysics of the NV center under infrared illumination, which is critical for many of the proposed exciting applications, and recently emerging fundamental research on optical trapping of diamond crystals. Clearly, varying the wavelength of the infrared laser, using excitation wavelengths that predominantly excite only one charge state (e.g. 593 nm laser), as well as performing statistical studies on single NV centers could be potentially useful future directions to investigate this effect. Another fruitful area might be to measure the electron capture and ionization rates ($r^{-/0}$ and $r^{0/-}$ in our paper) as a function of temperature, and apply principles of detailed balance to try and measure the energy separation between the charge states. 

\section{Acknowledgment}
This work was supported by the DOE Office of Basic Energy Sciences (DE-SC 0006638).
We thank Dr. Wolfgang J. Choyke and Dr. Robert P. Devaty for experimental support on indium soldering method and theoretical discussion.

\bibliography{ref}

\begin{thebibliography}{23}%
\makeatletter
\providecommand \@ifxundefined [1]{%
 \@ifx{#1\undefined}
}%
\providecommand \@ifnum [1]{%
 \ifnum #1\expandafter \@firstoftwo
 \else \expandafter \@secondoftwo
 \fi
}%
\providecommand \@ifx [1]{%
 \ifx #1\expandafter \@firstoftwo
 \else \expandafter \@secondoftwo
 \fi
}%
\providecommand \natexlab [1]{#1}%
\providecommand \enquote  [1]{``#1''}%
\providecommand \bibnamefont  [1]{#1}%
\providecommand \bibfnamefont [1]{#1}%
\providecommand \citenamefont [1]{#1}%
\providecommand \href@noop [0]{\@secondoftwo}%
\providecommand \href [0]{\begingroup \@sanitize@url \@href}%
\providecommand \@href[1]{\@@startlink{#1}\@@href}%
\providecommand \@@href[1]{\endgroup#1\@@endlink}%
\providecommand \@sanitize@url [0]{\catcode `\\12\catcode `\$12\catcode
  `\&12\catcode `\#12\catcode `\^12\catcode `\_12\catcode `\%12\relax}%
\providecommand \@@startlink[1]{}%
\providecommand \@@endlink[0]{}%
\providecommand \url  [0]{\begingroup\@sanitize@url \@url }%
\providecommand \@url [1]{\endgroup\@href {#1}{\urlprefix }}%
\providecommand \urlprefix  [0]{URL }%
\providecommand \Eprint [0]{\href }%
\providecommand \doibase [0]{http://dx.doi.org/}%
\providecommand \selectlanguage [0]{\@gobble}%
\providecommand \bibinfo  [0]{\@secondoftwo}%
\providecommand \bibfield  [0]{\@secondoftwo}%
\providecommand \translation [1]{[#1]}%
\providecommand \BibitemOpen [0]{}%
\providecommand \bibitemStop [0]{}%
\providecommand \bibitemNoStop [0]{.\EOS\space}%
\providecommand \EOS [0]{\spacefactor3000\relax}%
\providecommand \BibitemShut  [1]{\csname bibitem#1\endcsname}%
\let\auto@bib@innerbib\@empty
\bibitem [{\citenamefont {Doherty}\ \emph {et~al.}(2013)\citenamefont
  {Doherty}, \citenamefont {Manson}, \citenamefont {Delaney}, \citenamefont
  {Jelezko}, \citenamefont {Wrachtrup},\ and\ \citenamefont
  {Hollenberg}}]{Doherty13}%
  \BibitemOpen
  \bibfield  {author} {\bibinfo {author} {\bibfnamefont {M.~W.}\ \bibnamefont
  {Doherty}}, \bibinfo {author} {\bibfnamefont {N.~B.}\ \bibnamefont {Manson}},
  \bibinfo {author} {\bibfnamefont {P.}~\bibnamefont {Delaney}}, \bibinfo
  {author} {\bibfnamefont {F.}~\bibnamefont {Jelezko}}, \bibinfo {author}
  {\bibfnamefont {J.}~\bibnamefont {Wrachtrup}}, \ and\ \bibinfo {author}
  {\bibfnamefont {L.~C.}\ \bibnamefont {Hollenberg}},\ }\href {\doibase
  http://dx.doi.org/10.1016/j.physrep.2013.02.001} {\bibfield  {journal}
  {\bibinfo  {journal} {Physics Reports}\ }\textbf {\bibinfo {volume} {528}},\
  \bibinfo {pages} {1 } (\bibinfo {year} {2013})},\ \bibinfo {note} {the
  nitrogen-vacancy colour centre in diamond}\BibitemShut {NoStop}%
\bibitem [{\citenamefont {Schirhagl}\ \emph {et~al.}(2014)\citenamefont
  {Schirhagl}, \citenamefont {Chang}, \citenamefont {Loretz},\ and\
  \citenamefont {Degen}}]{Schirhagl14}%
  \BibitemOpen
  \bibfield  {author} {\bibinfo {author} {\bibfnamefont {R.}~\bibnamefont
  {Schirhagl}}, \bibinfo {author} {\bibfnamefont {K.}~\bibnamefont {Chang}},
  \bibinfo {author} {\bibfnamefont {M.}~\bibnamefont {Loretz}}, \ and\ \bibinfo
  {author} {\bibfnamefont {C.~L.}\ \bibnamefont {Degen}},\ }\href {\doibase
  10.1146/annurev-physchem-040513-103659} {\bibfield  {journal} {\bibinfo
  {journal} {Annual Review of Physical Chemistry}\ }\textbf {\bibinfo {volume}
  {65}},\ \bibinfo {pages} {83} (\bibinfo {year} {2014})},\ \bibinfo {note}
  {pMID: 24274702},\ \Eprint
  {http://arxiv.org/abs/http://dx.doi.org/10.1146/annurev-physchem-040513-103659}
  {http://dx.doi.org/10.1146/annurev-physchem-040513-103659} \BibitemShut
  {NoStop}%
\bibitem [{\citenamefont {Aslam}\ \emph {et~al.}(2013)\citenamefont {Aslam},
  \citenamefont {Waldherr}, \citenamefont {Neumann}, \citenamefont {Jelezko},\
  and\ \citenamefont {Wrachtrup}}]{Aslam13}%
  \BibitemOpen
  \bibfield  {author} {\bibinfo {author} {\bibfnamefont {N.}~\bibnamefont
  {Aslam}}, \bibinfo {author} {\bibfnamefont {G.}~\bibnamefont {Waldherr}},
  \bibinfo {author} {\bibfnamefont {P.}~\bibnamefont {Neumann}}, \bibinfo
  {author} {\bibfnamefont {F.}~\bibnamefont {Jelezko}}, \ and\ \bibinfo
  {author} {\bibfnamefont {J.}~\bibnamefont {Wrachtrup}},\ }\href
  {http://stacks.iop.org/1367-2630/15/i=1/a=013064} {\bibfield  {journal}
  {\bibinfo  {journal} {New Journal of Physics}\ }\textbf {\bibinfo {volume}
  {15}},\ \bibinfo {pages} {013064} (\bibinfo {year} {2013})}\BibitemShut
  {NoStop}%
\bibitem [{\citenamefont {Siyushev}\ \emph {et~al.}(2013)\citenamefont
  {Siyushev}, \citenamefont {Pinto}, \citenamefont {V\"or\"os}, \citenamefont
  {Gali}, \citenamefont {Jelezko},\ and\ \citenamefont
  {Wrachtrup}}]{Siyushev13}%
  \BibitemOpen
  \bibfield  {author} {\bibinfo {author} {\bibfnamefont {P.}~\bibnamefont
  {Siyushev}}, \bibinfo {author} {\bibfnamefont {H.}~\bibnamefont {Pinto}},
  \bibinfo {author} {\bibfnamefont {M.}~\bibnamefont {V\"or\"os}}, \bibinfo
  {author} {\bibfnamefont {A.}~\bibnamefont {Gali}}, \bibinfo {author}
  {\bibfnamefont {F.}~\bibnamefont {Jelezko}}, \ and\ \bibinfo {author}
  {\bibfnamefont {J.}~\bibnamefont {Wrachtrup}},\ }\href {\doibase
  10.1103/PhysRevLett.110.167402} {\bibfield  {journal} {\bibinfo  {journal}
  {Phys. Rev. Lett.}\ }\textbf {\bibinfo {volume} {110}},\ \bibinfo {pages}
  {167402} (\bibinfo {year} {2013})}\BibitemShut {NoStop}%
\bibitem [{\citenamefont {Shields}\ \emph {et~al.}(2015)\citenamefont
  {Shields}, \citenamefont {Unterreithmeier}, \citenamefont {de~Leon},
  \citenamefont {Park},\ and\ \citenamefont {Lukin}}]{Shields15}%
  \BibitemOpen
  \bibfield  {author} {\bibinfo {author} {\bibfnamefont {B.~J.}\ \bibnamefont
  {Shields}}, \bibinfo {author} {\bibfnamefont {Q.~P.}\ \bibnamefont
  {Unterreithmeier}}, \bibinfo {author} {\bibfnamefont {N.~P.}\ \bibnamefont
  {de~Leon}}, \bibinfo {author} {\bibfnamefont {H.}~\bibnamefont {Park}}, \
  and\ \bibinfo {author} {\bibfnamefont {M.~D.}\ \bibnamefont {Lukin}},\ }\href
  {\doibase 10.1103/PhysRevLett.114.136402} {\bibfield  {journal} {\bibinfo
  {journal} {Phys. Rev. Lett.}\ }\textbf {\bibinfo {volume} {114}},\ \bibinfo
  {pages} {136402} (\bibinfo {year} {2015})}\BibitemShut {NoStop}%
\bibitem [{\citenamefont {Yin}\ \emph {et~al.}(2013)\citenamefont {Yin},
  \citenamefont {Li}, \citenamefont {Zhang},\ and\ \citenamefont
  {Duan}}]{Yin13}%
  \BibitemOpen
  \bibfield  {author} {\bibinfo {author} {\bibfnamefont {Z.~Q.}\ \bibnamefont
  {Yin}}, \bibinfo {author} {\bibfnamefont {T.}~\bibnamefont {Li}}, \bibinfo
  {author} {\bibfnamefont {X.}~\bibnamefont {Zhang}}, \ and\ \bibinfo {author}
  {\bibfnamefont {L.~M.}\ \bibnamefont {Duan}},\ }\href {\doibase
  10.1103/PhysRevA.88.033614} {\bibfield  {journal} {\bibinfo  {journal} {Phys.
  Rev. A}\ }\textbf {\bibinfo {volume} {88}},\ \bibinfo {pages} {033614}
  (\bibinfo {year} {2013})}\BibitemShut {NoStop}%
\bibitem [{\citenamefont {Neukirch}\ \emph {et~al.}(2013)\citenamefont
  {Neukirch}, \citenamefont {Gieseler}, \citenamefont {Quidant}, \citenamefont
  {Novotny},\ and\ \citenamefont {Vamivakas}}]{Neukirch13}%
  \BibitemOpen
  \bibfield  {author} {\bibinfo {author} {\bibfnamefont {L.~P.}\ \bibnamefont
  {Neukirch}}, \bibinfo {author} {\bibfnamefont {J.}~\bibnamefont {Gieseler}},
  \bibinfo {author} {\bibfnamefont {R.}~\bibnamefont {Quidant}}, \bibinfo
  {author} {\bibfnamefont {L.}~\bibnamefont {Novotny}}, \ and\ \bibinfo
  {author} {\bibfnamefont {A.~N.}\ \bibnamefont {Vamivakas}},\ }\href {\doibase
  10.1364/OL.38.002976} {\bibfield  {journal} {\bibinfo  {journal} {Opt.
  Lett.}\ }\textbf {\bibinfo {volume} {38}},\ \bibinfo {pages} {2976} (\bibinfo
  {year} {2013})}\BibitemShut {NoStop}%
\bibitem [{\citenamefont {Hoang}\ \emph {et~al.}(2015)\citenamefont {Hoang},
  \citenamefont {Ahn}, \citenamefont {Bang},\ and\ \citenamefont
  {Li}}]{Hoang15}%
  \BibitemOpen
  \bibfield  {author} {\bibinfo {author} {\bibfnamefont {T.~M.}\ \bibnamefont
  {Hoang}}, \bibinfo {author} {\bibfnamefont {J.}~\bibnamefont {Ahn}}, \bibinfo
  {author} {\bibfnamefont {J.}~\bibnamefont {Bang}}, \ and\ \bibinfo {author}
  {\bibfnamefont {T.}~\bibnamefont {Li}},\ }\href@noop {} {\bibfield  {journal}
  {\bibinfo  {journal} {arXiv:1510.06715}\ } (\bibinfo {year}
  {2015})}\BibitemShut {NoStop}%
\bibitem [{\citenamefont {Lai}\ \emph {et~al.}(2013)\citenamefont {Lai},
  \citenamefont {Faklaris}, \citenamefont {Zheng}, \citenamefont {Jacques},
  \citenamefont {Chang}, \citenamefont {Roch},\ and\ \citenamefont
  {Treussart}}]{Lai13}%
  \BibitemOpen
  \bibfield  {author} {\bibinfo {author} {\bibfnamefont {N.~D.}\ \bibnamefont
  {Lai}}, \bibinfo {author} {\bibfnamefont {O.}~\bibnamefont {Faklaris}},
  \bibinfo {author} {\bibfnamefont {D.}~\bibnamefont {Zheng}}, \bibinfo
  {author} {\bibfnamefont {V.}~\bibnamefont {Jacques}}, \bibinfo {author}
  {\bibfnamefont {H.-C.}\ \bibnamefont {Chang}}, \bibinfo {author}
  {\bibfnamefont {J.-F.}\ \bibnamefont {Roch}}, \ and\ \bibinfo {author}
  {\bibfnamefont {F.}~\bibnamefont {Treussart}},\ }\href
  {http://stacks.iop.org/1367-2630/15/i=3/a=033030} {\bibfield  {journal}
  {\bibinfo  {journal} {New Journal of Physics}\ }\textbf {\bibinfo {volume}
  {15}},\ \bibinfo {pages} {033030} (\bibinfo {year} {2013})}\BibitemShut
  {NoStop}%
\bibitem [{\citenamefont {Geiselmann}\ \emph {et~al.}(2013)\citenamefont
  {Geiselmann}, \citenamefont {Marty}, \citenamefont {Garcia~de Abajo},\ and\
  \citenamefont {Quidant}}]{Geiselmann13}%
  \BibitemOpen
  \bibfield  {author} {\bibinfo {author} {\bibfnamefont {M.}~\bibnamefont
  {Geiselmann}}, \bibinfo {author} {\bibfnamefont {R.}~\bibnamefont {Marty}},
  \bibinfo {author} {\bibfnamefont {F.~J.}\ \bibnamefont {Garcia~de Abajo}}, \
  and\ \bibinfo {author} {\bibfnamefont {R.}~\bibnamefont {Quidant}},\ }\href
  {http://dx.doi.org/10.1038/nphys2770} {\bibfield  {journal} {\bibinfo
  {journal} {Nat Phys}\ }\textbf {\bibinfo {volume} {9}},\ \bibinfo {pages}
  {785} (\bibinfo {year} {2013})}\BibitemShut {NoStop}%
\bibitem [{\citenamefont {Gaebel}\ \emph {et~al.}(2005)\citenamefont {Gaebel},
  \citenamefont {Domhan}, \citenamefont {Wittmann}, \citenamefont {Popa},
  \citenamefont {Jelezko}, \citenamefont {Rabeau}, \citenamefont {Greentree},
  \citenamefont {Prawer}, \citenamefont {Trajkov}, \citenamefont {Hemmer},\
  and\ \citenamefont {Wrachtrup}}]{Gaebel05}%
  \BibitemOpen
  \bibfield  {author} {\bibinfo {author} {\bibfnamefont {T.}~\bibnamefont
  {Gaebel}}, \bibinfo {author} {\bibfnamefont {M.}~\bibnamefont {Domhan}},
  \bibinfo {author} {\bibfnamefont {C.}~\bibnamefont {Wittmann}}, \bibinfo
  {author} {\bibfnamefont {I.}~\bibnamefont {Popa}}, \bibinfo {author}
  {\bibfnamefont {F.}~\bibnamefont {Jelezko}}, \bibinfo {author} {\bibfnamefont
  {J.}~\bibnamefont {Rabeau}}, \bibinfo {author} {\bibfnamefont
  {A.}~\bibnamefont {Greentree}}, \bibinfo {author} {\bibfnamefont
  {S.}~\bibnamefont {Prawer}}, \bibinfo {author} {\bibfnamefont
  {E.}~\bibnamefont {Trajkov}}, \bibinfo {author} {\bibfnamefont
  {P.}~\bibnamefont {Hemmer}}, \ and\ \bibinfo {author} {\bibfnamefont
  {J.}~\bibnamefont {Wrachtrup}},\ }\href {\doibase 10.1007/s00340-005-2056-2}
  {\bibfield  {journal} {\bibinfo  {journal} {Applied Physics B}\ }\textbf
  {\bibinfo {volume} {82}},\ \bibinfo {pages} {243} (\bibinfo {year}
  {2005})}\BibitemShut {NoStop}%
\bibitem [{\citenamefont {Rondin}\ \emph {et~al.}(2010)\citenamefont {Rondin},
  \citenamefont {Dantelle}, \citenamefont {Slablab}, \citenamefont {Grosshans},
  \citenamefont {Treussart}, \citenamefont {Bergonzo}, \citenamefont
  {Perruchas}, \citenamefont {Gacoin}, \citenamefont {Chaigneau}, \citenamefont
  {Chang}, \citenamefont {Jacques},\ and\ \citenamefont {Roch}}]{Rondin10}%
  \BibitemOpen
  \bibfield  {author} {\bibinfo {author} {\bibfnamefont {L.}~\bibnamefont
  {Rondin}}, \bibinfo {author} {\bibfnamefont {G.}~\bibnamefont {Dantelle}},
  \bibinfo {author} {\bibfnamefont {A.}~\bibnamefont {Slablab}}, \bibinfo
  {author} {\bibfnamefont {F.}~\bibnamefont {Grosshans}}, \bibinfo {author}
  {\bibfnamefont {F.}~\bibnamefont {Treussart}}, \bibinfo {author}
  {\bibfnamefont {P.}~\bibnamefont {Bergonzo}}, \bibinfo {author}
  {\bibfnamefont {S.}~\bibnamefont {Perruchas}}, \bibinfo {author}
  {\bibfnamefont {T.}~\bibnamefont {Gacoin}}, \bibinfo {author} {\bibfnamefont
  {M.}~\bibnamefont {Chaigneau}}, \bibinfo {author} {\bibfnamefont {H.-C.}\
  \bibnamefont {Chang}}, \bibinfo {author} {\bibfnamefont {V.}~\bibnamefont
  {Jacques}}, \ and\ \bibinfo {author} {\bibfnamefont {J.-F.}\ \bibnamefont
  {Roch}},\ }\href {\doibase 10.1103/PhysRevB.82.115449} {\bibfield  {journal}
  {\bibinfo  {journal} {Phys. Rev. B}\ }\textbf {\bibinfo {volume} {82}},\
  \bibinfo {pages} {115449} (\bibinfo {year} {2010})}\BibitemShut {NoStop}%
\bibitem [{\citenamefont {Savitzky}\ and\ \citenamefont
  {Golay}(1964)}]{savitzky1964}%
  \BibitemOpen
  \bibfield  {author} {\bibinfo {author} {\bibfnamefont {A.}~\bibnamefont
  {Savitzky}}\ and\ \bibinfo {author} {\bibfnamefont {M.~J.~E.}\ \bibnamefont
  {Golay}},\ }\href {\doibase 10.1021/ac60214a047} {\bibfield  {journal}
  {\bibinfo  {journal} {Analytical Chemistry}\ }\textbf {\bibinfo {volume}
  {36}},\ \bibinfo {pages} {1627} (\bibinfo {year} {1964})},\ \Eprint
  {http://arxiv.org/abs/http://dx.doi.org/10.1021/ac60214a047}
  {http://dx.doi.org/10.1021/ac60214a047} \BibitemShut {NoStop}%
\bibitem [{\citenamefont {Manson}\ and\ \citenamefont
  {Harrison}(2005)}]{Manson05}%
  \BibitemOpen
  \bibfield  {author} {\bibinfo {author} {\bibfnamefont {N.}~\bibnamefont
  {Manson}}\ and\ \bibinfo {author} {\bibfnamefont {J.}~\bibnamefont
  {Harrison}},\ }\href {\doibase
  http://dx.doi.org/10.1016/j.diamond.2005.06.027} {\bibfield  {journal}
  {\bibinfo  {journal} {Diamond and Related Materials}\ }\textbf {\bibinfo
  {volume} {14}},\ \bibinfo {pages} {1705 } (\bibinfo {year}
  {2005})}\BibitemShut {NoStop}%
\bibitem [{\citenamefont {Waldherr}\ \emph {et~al.}(2011)\citenamefont
  {Waldherr}, \citenamefont {Beck}, \citenamefont {Steiner}, \citenamefont
  {Neumann}, \citenamefont {Gali}, \citenamefont {Frauenheim}, \citenamefont
  {Jelezko},\ and\ \citenamefont {Wrachtrup}}]{Waldherr11}%
  \BibitemOpen
  \bibfield  {author} {\bibinfo {author} {\bibfnamefont {G.}~\bibnamefont
  {Waldherr}}, \bibinfo {author} {\bibfnamefont {J.}~\bibnamefont {Beck}},
  \bibinfo {author} {\bibfnamefont {M.}~\bibnamefont {Steiner}}, \bibinfo
  {author} {\bibfnamefont {P.}~\bibnamefont {Neumann}}, \bibinfo {author}
  {\bibfnamefont {A.}~\bibnamefont {Gali}}, \bibinfo {author} {\bibfnamefont
  {T.}~\bibnamefont {Frauenheim}}, \bibinfo {author} {\bibfnamefont
  {F.}~\bibnamefont {Jelezko}}, \ and\ \bibinfo {author} {\bibfnamefont
  {J.}~\bibnamefont {Wrachtrup}},\ }\href {\doibase
  10.1103/PhysRevLett.106.157601} {\bibfield  {journal} {\bibinfo  {journal}
  {Phys. Rev. Lett.}\ }\textbf {\bibinfo {volume} {106}},\ \bibinfo {pages}
  {157601} (\bibinfo {year} {2011})}\BibitemShut {NoStop}%
\bibitem [{\citenamefont {Beha}\ \emph {et~al.}(2012)\citenamefont {Beha},
  \citenamefont {Batalov}, \citenamefont {Manson}, \citenamefont
  {Bratschitsch},\ and\ \citenamefont {Leitenstorfer}}]{Beha12}%
  \BibitemOpen
  \bibfield  {author} {\bibinfo {author} {\bibfnamefont {K.}~\bibnamefont
  {Beha}}, \bibinfo {author} {\bibfnamefont {A.}~\bibnamefont {Batalov}},
  \bibinfo {author} {\bibfnamefont {N.~B.}\ \bibnamefont {Manson}}, \bibinfo
  {author} {\bibfnamefont {R.}~\bibnamefont {Bratschitsch}}, \ and\ \bibinfo
  {author} {\bibfnamefont {A.}~\bibnamefont {Leitenstorfer}},\ }\href {\doibase
  10.1103/PhysRevLett.109.097404} {\bibfield  {journal} {\bibinfo  {journal}
  {Phys. Rev. Lett.}\ }\textbf {\bibinfo {volume} {109}},\ \bibinfo {pages}
  {097404} (\bibinfo {year} {2012})}\BibitemShut {NoStop}%
\bibitem [{\citenamefont {Robledo}\ \emph {et~al.}(2010)\citenamefont
  {Robledo}, \citenamefont {Bernien}, \citenamefont {van Weperen},\ and\
  \citenamefont {Hanson}}]{Robledo10}%
  \BibitemOpen
  \bibfield  {author} {\bibinfo {author} {\bibfnamefont {L.}~\bibnamefont
  {Robledo}}, \bibinfo {author} {\bibfnamefont {H.}~\bibnamefont {Bernien}},
  \bibinfo {author} {\bibfnamefont {I.}~\bibnamefont {van Weperen}}, \ and\
  \bibinfo {author} {\bibfnamefont {R.}~\bibnamefont {Hanson}},\ }\href
  {\doibase 10.1103/PhysRevLett.105.177403} {\bibfield  {journal} {\bibinfo
  {journal} {Phys. Rev. Lett.}\ }\textbf {\bibinfo {volume} {105}},\ \bibinfo
  {pages} {177403} (\bibinfo {year} {2010})}\BibitemShut {NoStop}%
\bibitem [{\citenamefont {Walker}(1979)}]{Walker1979}%
  \BibitemOpen
  \bibfield  {author} {\bibinfo {author} {\bibfnamefont {J.}~\bibnamefont
  {Walker}},\ }\href {http://stacks.iop.org/0034-4885/42/i=10/a=001} {\bibfield
   {journal} {\bibinfo  {journal} {Reports on Progress in Physics}\ }\textbf
  {\bibinfo {volume} {42}},\ \bibinfo {pages} {1605} (\bibinfo {year}
  {1979})}\BibitemShut {NoStop}%
\bibitem [{\citenamefont {Jeske}\ \emph {et~al.}(2016)\citenamefont {Jeske},
  \citenamefont {Lau}, \citenamefont {McGuinness}, \citenamefont {Reineck},
  \citenamefont {Johnson}, \citenamefont {McCallum}, \citenamefont {Jelezko},
  \citenamefont {Volz}, \citenamefont {Cole}, \citenamefont {Gibson},\ and\
  \citenamefont {Greentree}}]{Jeske16}%
  \BibitemOpen
  \bibfield  {author} {\bibinfo {author} {\bibfnamefont {J.}~\bibnamefont
  {Jeske}}, \bibinfo {author} {\bibfnamefont {D.~W.~M.}\ \bibnamefont {Lau}},
  \bibinfo {author} {\bibfnamefont {L.~P.}\ \bibnamefont {McGuinness}},
  \bibinfo {author} {\bibfnamefont {P.}~\bibnamefont {Reineck}}, \bibinfo
  {author} {\bibfnamefont {B.~C.}\ \bibnamefont {Johnson}}, \bibinfo {author}
  {\bibfnamefont {J.~C.}\ \bibnamefont {McCallum}}, \bibinfo {author}
  {\bibfnamefont {F.}~\bibnamefont {Jelezko}}, \bibinfo {author} {\bibfnamefont
  {T.}~\bibnamefont {Volz}}, \bibinfo {author} {\bibfnamefont {J.~H.}\
  \bibnamefont {Cole}}, \bibinfo {author} {\bibfnamefont {B.~C.}\ \bibnamefont
  {Gibson}}, \ and\ \bibinfo {author} {\bibfnamefont {A.~D.}\ \bibnamefont
  {Greentree}},\ }\href@noop {} {\bibfield  {journal} {\bibinfo  {journal}
  {arXiv:1602.07418}\ } (\bibinfo {year} {2016})}\BibitemShut {NoStop}%
\bibitem [{\citenamefont {Acosta}\ \emph {et~al.}(2010)\citenamefont {Acosta},
  \citenamefont {Bauch}, \citenamefont {Ledbetter}, \citenamefont {Waxman},
  \citenamefont {Bouchard},\ and\ \citenamefont {Budker}}]{Acosta10}%
  \BibitemOpen
  \bibfield  {author} {\bibinfo {author} {\bibfnamefont {V.~M.}\ \bibnamefont
  {Acosta}}, \bibinfo {author} {\bibfnamefont {E.}~\bibnamefont {Bauch}},
  \bibinfo {author} {\bibfnamefont {M.~P.}\ \bibnamefont {Ledbetter}}, \bibinfo
  {author} {\bibfnamefont {A.}~\bibnamefont {Waxman}}, \bibinfo {author}
  {\bibfnamefont {L.-S.}\ \bibnamefont {Bouchard}}, \ and\ \bibinfo {author}
  {\bibfnamefont {D.}~\bibnamefont {Budker}},\ }\href {\doibase
  10.1103/PhysRevLett.104.070801} {\bibfield  {journal} {\bibinfo  {journal}
  {Phys. Rev. Lett.}\ }\textbf {\bibinfo {volume} {104}},\ \bibinfo {pages}
  {070801} (\bibinfo {year} {2010})}\BibitemShut {NoStop}%
\bibitem [{\citenamefont {Kucsko}\ \emph {et~al.}(2013)\citenamefont {Kucsko},
  \citenamefont {Maurer}, \citenamefont {Yao}, \citenamefont {Kubo},
  \citenamefont {Noh}, \citenamefont {Lo}, \citenamefont {Park},\ and\
  \citenamefont {Lukin}}]{Kucsko13}%
  \BibitemOpen
  \bibfield  {author} {\bibinfo {author} {\bibfnamefont {G.}~\bibnamefont
  {Kucsko}}, \bibinfo {author} {\bibfnamefont {P.~C.}\ \bibnamefont {Maurer}},
  \bibinfo {author} {\bibfnamefont {N.~Y.}\ \bibnamefont {Yao}}, \bibinfo
  {author} {\bibfnamefont {M.}~\bibnamefont {Kubo}}, \bibinfo {author}
  {\bibfnamefont {H.~J.}\ \bibnamefont {Noh}}, \bibinfo {author} {\bibfnamefont
  {P.~K.}\ \bibnamefont {Lo}}, \bibinfo {author} {\bibfnamefont
  {H.}~\bibnamefont {Park}}, \ and\ \bibinfo {author} {\bibfnamefont {M.~D.}\
  \bibnamefont {Lukin}},\ }\href {http://dx.doi.org/10.1038/nature12373}
  {\bibfield  {journal} {\bibinfo  {journal} {Nature}\ }\textbf {\bibinfo
  {volume} {500}},\ \bibinfo {pages} {54} (\bibinfo {year} {2013})},\ \bibinfo
  {note} {letter}\BibitemShut {NoStop}%
\bibitem [{\citenamefont {Jelezko}\ \emph {et~al.}(2004)\citenamefont
  {Jelezko}, \citenamefont {Gaebel}, \citenamefont {Popa}, \citenamefont
  {Gruber},\ and\ \citenamefont {Wrachtrup}}]{Jelezko04}%
  \BibitemOpen
  \bibfield  {author} {\bibinfo {author} {\bibfnamefont {F.}~\bibnamefont
  {Jelezko}}, \bibinfo {author} {\bibfnamefont {T.}~\bibnamefont {Gaebel}},
  \bibinfo {author} {\bibfnamefont {I.}~\bibnamefont {Popa}}, \bibinfo {author}
  {\bibfnamefont {A.}~\bibnamefont {Gruber}}, \ and\ \bibinfo {author}
  {\bibfnamefont {J.}~\bibnamefont {Wrachtrup}},\ }\href {\doibase
  10.1103/PhysRevLett.92.076401} {\bibfield  {journal} {\bibinfo  {journal}
  {Phys. Rev. Lett.}\ }\textbf {\bibinfo {volume} {92}},\ \bibinfo {pages}
  {076401} (\bibinfo {year} {2004})}\BibitemShut {NoStop}%
\bibitem [{\citenamefont {Childress}\ \emph {et~al.}(2006)\citenamefont
  {Childress}, \citenamefont {Gurudev~Dutt}, \citenamefont {Taylor},
  \citenamefont {Zibrov}, \citenamefont {Jelezko}, \citenamefont {Wrachtrup},
  \citenamefont {Hemmer},\ and\ \citenamefont {Lukin}}]{Childress06}%
  \BibitemOpen
  \bibfield  {author} {\bibinfo {author} {\bibfnamefont {L.}~\bibnamefont
  {Childress}}, \bibinfo {author} {\bibfnamefont {M.~V.}\ \bibnamefont
  {Gurudev~Dutt}}, \bibinfo {author} {\bibfnamefont {J.~M.}\ \bibnamefont
  {Taylor}}, \bibinfo {author} {\bibfnamefont {A.~S.}\ \bibnamefont {Zibrov}},
  \bibinfo {author} {\bibfnamefont {F.}~\bibnamefont {Jelezko}}, \bibinfo
  {author} {\bibfnamefont {J.}~\bibnamefont {Wrachtrup}}, \bibinfo {author}
  {\bibfnamefont {P.~R.}\ \bibnamefont {Hemmer}}, \ and\ \bibinfo {author}
  {\bibfnamefont {M.~D.}\ \bibnamefont {Lukin}},\ }\href {\doibase
  10.1126/science.1131871} {\bibfield  {journal} {\bibinfo  {journal}
  {Science}\ }\textbf {\bibinfo {volume} {314}},\ \bibinfo {pages} {281}
  (\bibinfo {year} {2006})}\BibitemShut {NoStop}%
\end{thebibliography}%
\end{document}